\documentclass[10pt]{elsart5p}
\usepackage{amsmath}
\usepackage{amssymb}
\usepackage{latexsym}
\usepackage{bbm}
\usepackage{graphics}
\usepackage{float}
\usepackage{times}
\newcommand{\DD}{\mathit{\Delta}}
\begin{document}
\begin{frontmatter}
\title{Quantum Coherence Conservation by Growth in
Environmental Dissipation Rate}
\author[Osaka]{Akira SaiToh\corauthref{cor}}
\ead{saitoh@qc.ee.es.osaka-u.ac.jp}
\corauth[cor]{Corresponding author.}
\author[Kinki]{Robabeh Rahimi}
\ead{rahimi@alice.math.kindai.ac.jp}
\author[Kinki]{Mikio Nakahara}
\ead{nakahara@math.kindai.ac.jp}
\address[Osaka]{Department of Systems Innovation, Graduate
School of Engineering Science, Osaka University, Toyonaka, Osaka
560-8531, Japan}
\address[Kinki]{Department of Physics, Kinki University, Higashi-Osaka,
Osaka 577-8502, Japan}
\date{uploaded on October 23, 2007}
\begin{abstract}
Quantum coherence conservation is shown to be achieved by a very high
rate of dissipation of an environmental system coupled with a principal
system. This effect is not in the list of previously-known strategies of
noise suppression, such as Zeno effect, dynamical decoupling, quantum
error correction code, and decoherence free subspace. An analytical
solution is found for a simplified model of a single qubit coupled with
an environmental single qubit dissipating rapidly. We also show examples
of coherence conservation in a spin-boson linear coupling model with a
numerical evaluation.
\end{abstract}
%
\end{frontmatter}
Suppressing decoherence in quantum systems is of growing interest
in the light of the rapid development of quantum computing \cite{G99,NC00}.
There have been many schemes proposed for this purpose in several
categories, namely, Zeno effect \cite{Zeno}, dynamical decoupling
(or bang-bang control) \cite{BB98,VL98}, quantum error correction
code (QECC) \cite{S95,G96,L96}, and decoherence free subspace (DFS)
\cite{ZR97,DG97,LCW98}.
Conventional schemes for suppressing decoherence focus on controlling
a principal system that is under influence of environmental systems.
The quantum Zeno effect uses a sequence of projective operations
(usually projective measurements) mapping states onto 
subspaces of a state space \cite{FP02}. A quantum state is kept
inside of a subspace if projective operations are applied very
frequently before noise kicks the state out of the subspace.
Dynamical decoupling has been studied intensively in the field of
nuclear magnetic resonance (NMR) from the dawn of the field (see, e.g.,
Ref.\ \cite{Ernst}) and later extended to other physical systems
\cite{BB98,VL98,VT99,WL04,SHHMM05}. This scheme uses a train of regular
short pulses applied to a principal system to cancel time evolutions
under noise. 
The concept of QECC is to discard a subspace easily affected by noise.
A QECC uses a space of code words generated from original states by
adding a certain redundancy so that a recovery from corruption is possible.
The DFS scheme utilizes a subspace of states that are unchanged by given
noise operators. Quantum computation is performed in the subspace.

Studies on coherence conservation so far are mostly based on the
assumption that we do not have control over environment. There may be
alternatives available if we focus on the idea of controlling
environmental parameters. A method to control effective coupling using a
dynamical control field in the presence of time-dependent external field
was recently reported by Jirari and P\"otz \cite{JP06,JP07}. We pursue
the possibility of passive control, rather than dynamical control, in a
simple picture of decoherence (See Ref.\ \cite{Louisell} for
conventional decoherence models).

There are many parameters affecting the decoherence factor in realistic
models. It is often mentioned that there are the strong coupling regime
and the weak coupling regime for a system consisting of a principal
system and a noise source in general. (A typical case is a qubit under
a random telegraphic noise \cite{SCCD05}.) The two regimes involve
significantly different dependencies of the decoherence factor on the
parameters describing noise. We need to choose a proper parameter with
which one can alter the behaviour of the decoherence factor for our purpose.

We report in this letter coherence conservation achieved by a very
high rate of dissipation of an environmental system coupled with a
principal system. A highly dissipative environmental system is wiped
out before affecting the principal system.
A clue of the phenomenon was originally found in our
numerical simulation of bang-bang control of entanglement in a spin-bus
model \cite{RSN}. We investigate the phenomenon in detail by using a
mathematical analysis and a numerical simulation for simplified models.
An analytical solution is found for a model of a single qubit coupled
with an environmental single qubit dissipating rapidly. The effect is
also found in a spin-boson linear coupling model by using a numerical
evaluation.

\begin{figure}[tb]
\begin{center}
 \scalebox{0.6}{\includegraphics{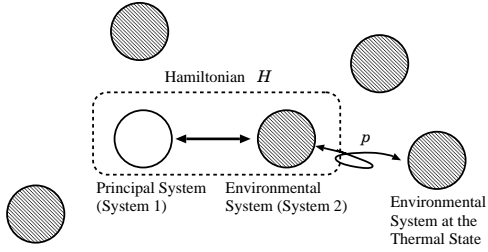}}
\caption{\label{figsimplemodel}
Model for the system consisting of the principal system (system 1)
and the environmental system (system 2) whose time evolution is
governed by the Hamiltonian $H$. System 2 is replaced with a
thermal environmental system with the dissipation probability $p$
for the time interval $\tau$.}
\end{center}
\end{figure}

Let us assume that the entire environment is so large that the environmental
system (system 2) coupled with the principal system (system 1) is a part of a
large environment and hence it is replaced with a thermal environmental
system with probability $p$ (namely, with some dissipation rate)
per certain time interval $\tau$. Systems 1 and 2 are represented by
the density matrix $\rho^{[1,2]}$; a thermal environmental system is
represented by the density matrix $\sigma$.
The Hamiltonian affecting the time evolution is reduced to the one
consisting only of the time-independent Hamiltonian $H$ that governs
systems 1 and 2 including their interaction. This model is illustrated in
Fig.\ \ref{figsimplemodel}. For a small time interval $\DD t$, the
evolution of the systems 1 and 2 obeys the equation
\begin{equation}\label{eqMain}\begin{split}
\rho^{[1,2]}(\tilde{t}+&\DD t)=e^{-iH\DD t}\biggl[
x^{\DD t}\rho^{[1,2]}(\tilde{t})\\
&+(1-x^{\DD t}){\rm Tr}_2\rho^{[1,2]}(\tilde{t})
\otimes \sigma\biggr]e^{iH\DD t},
\end{split}\end{equation}
where $x=(1-p)^{1/\tau}$ and $\tilde{t}$ denotes a certain time step.
The dissipation rate $p$ can be modified by changing the experimental
setup under a static control. 


Let us begin with an analytical evaluation of this model in the case
where the systems 1 and 2 are single-qubit systems. For further simplicity,
we impose the following conditions. The principal system is originally
represented by a density matrix
\[
 \rho^{[1]}(0)=\begin{pmatrix}a&b\\b^*&1-a\end{pmatrix}
\]
with $0\le a \le 1$ and $0\le |b|\le \sqrt{a(1-a)}$.
The environmental system at thermal equilibrium is represented by the
maximally-mixed density matrix
\[
 \rho^{[2]}(0)=\sigma=\begin{pmatrix}1/2&0\\0&1/2\end{pmatrix}.
\]
The initial state of the total system is set to
$
 \rho^{[1,2]}(0)=\rho^{[1]}(0)\otimes \rho^{[2]}(0).
$
The Hamiltonian $H$ is set to
$cI_z\otimes I_z=\mathrm{diag}(c/4,-c/4,-c/4,c/4)$
[here, $I_z=\mathrm{diag}(1/2, -1/2)$].

With these simplifications, one can find the form of the density matrix
at time $t=m\DD t$ ($m\in\{0,1,2,\ldots\}$) as
\[\rho^{[1,2]}(m\DD t)=
 \begin{pmatrix}
a/2&0&f_m&0\\
0&a/2&0&g_m\\
f^*_m&0&(1-a)/2&0\\
0&g^*_m&0&(1-a)/2
\end{pmatrix},
\]
with functions $f_m$ and $g_m$ depending on $m$.
The functions $f_m$ and $g_m$ obey the system of recurrence formulae:
\[
\left\{
\begin{array}{l}
f_{m+1}=\frac{1}{2}e^{-ic\DD t/2}\left[
f_m+g_m+x^{\DD t}(f_m-g_m)
\right]\\
g_{m+1}=\frac{1}{2}e^{ic\DD t/2}\left[
f_m+g_m-x^{\DD t}(f_m-g_m)
\right]
\end{array}
\right.
\]
with $f_0=g_0=b/2$.
This leads to the following recurrence formula:
\begin{equation}\label{EqRec}
 \kappa_{m+2}=(1+x^{\DD t})\cos(c\DD t/2) \kappa_{m+1}-x^{\DD t}\kappa_m,
\end{equation}
where $\kappa_m=f_m$ or $g_m$ 
with $f_0=g_0=b/2$, $f_ 1=be^{-ic\DD t/2}/2$, and $g_1=be^{ic\DD t/2}/2$.

One can derive functions
$f(t)={\rm lim}_{\DD t\rightarrow 0, m\DD t = t}f_m$ and
$g(t)={\rm lim}_{\DD t\rightarrow 0, m\DD t = t}g_m$ in the following way.
By linearlization, Eq.\ (\ref{EqRec}) is put in the form:
\[\begin{split}
 &\kappa_{m+2}-2\kappa_{m+1}+\kappa_m
-\DD t\ln x(\kappa_{m+1}-\kappa_m)\\
&+(\DD t)^2\frac{c^2}{4}\kappa_{m+1}
-\frac{(\DD t)^2}{2}(\ln x)^2(\kappa_{m+1}-\kappa_m)\\
&+\mathcal{O}[(\DD t)^3]=0
\end{split}\]
Dividing this equation by $(\DD t)^2$ and taking the limit
$\DD t\rightarrow 0$ lead to
\[
 \partial^2\kappa(t)/\partial t^2 -\ln x ~\partial\kappa(t)/\partial t
+c^2\kappa(t)/4 =0,
\]
where $\kappa(t)={\rm lim}_{\DD t\rightarrow 0, m\DD t = t}\kappa_m$.
The solution of this differential equation is
\[
 \kappa(t)=u_\kappa e^{-r_+ t} + v_\kappa e^{-r_- t}
\]
with constants $u_\kappa$ and $v_\kappa$ ($\kappa=f$ or $g$),
and the complex decoherence factor
\begin{equation*}\label{eqr}
r_\pm=-\left[\ln x\pm\sqrt{(\ln x)^2-c^2}\right]/2.
\end{equation*}
The real part of $r_\pm$ is nonnegative because $\ln x = \ln (1-p)/\tau \le 0$
and $|\ln x| \ge |{\rm Re}\sqrt{(\ln x)^2-c^2}|$.
(Here, the square root is positive.)

We need to impose the conditions that 
$\kappa(0)= b/2$ and $\kappa'(0)=\lim_{\DD t\rightarrow 0}
(\kappa_1-\kappa_0)/\DD t$. The latter condition can be
written as $-r_+u_f -r_-v_f=-ibc/4$
and $-r_+u_g -r_-v_g=ibc/4$. With these conditions, we obtain
\[\begin{split}
u_f &= \frac{ibc-2br_-}{4(r_+-r_-)},~~
v_f =\frac{-ibc+2br_+}{4(r_+-r_-)},\\
u_g &= \frac{-ibc-2br_-}{4(r_+-r_-)},~~
v_g = \frac{ibc+2br_+}{4(r_+-r_-)}. 
\end{split}\]
Consequently, we have
\[
 f(t)=\frac{ibc-2br_-}{4(r_+-r_-)}e^{-r_+t}+
\frac{-ibc+2br_+}{4(r_+-r_-)}e^{-r_-t},
\]
\[
 g(t)=\frac{-ibc-2br_-}{4(r_+-r_-)}e^{-r_+t}+
\frac{ibc+2br_+}{4(r_+-r_-)}e^{-r_-t}.
\]

\begin{figure}[hpbt]\begin{center}
\scalebox{0.6}{\includegraphics{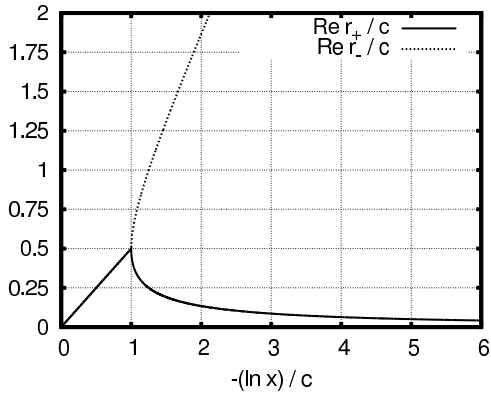}}
\caption{\label{figr}Plot of ${\rm Re}~r_\pm/c$ as functions of $-(\ln x) / c$.
The decoherence factors ${\rm Re}~r_\pm$ increases until they reaches
$c/2$ as the dissipation rate increases. The factor ${\rm Re}~r_+$ starts
decreasing at the point of $-\ln x = c$ (i.e., $p=1-e^{-c\tau}$) while the
factor ${\rm Re}~r_-$ starts increasing rapidly at this point.}
\end{center}\end{figure}

One can now write the reduced density matrix of the principal system at
$t$ as
\begin{equation*}\label{EqFinal}
 \rho^{[1]}(t)=\begin{pmatrix}
a& \eta(t)\\
\eta(t)^*&1-a
\end{pmatrix}
\end{equation*}
with
\[
 \eta(t)=b\left(\frac{-r_-}{r_+-r_-}e^{-r_+ t}
               +\frac{r_+}{r_+-r_-}e^{-r_- t}\right).
\]
Let us investigate $r_\pm$ in details in relation to $p$. One obvious fact
is that exponential decay is caused by the real part of $e^{-r_\pm t}$.
Thus we focus on the behaviour of ${\rm Re}~r_\pm$. We find that
${\rm Re}~r_+$ increases as $-(\ln x)/2$ when $0\le p\le 1-e^{-c\tau}$
and decreases with convergence to zero when $1-e^{-c\tau}< p \le 1$.
In contrast, ${\rm Re}~r_-$ increases as $-(\ln x)/2$ when $0\le p\le
1-e^{-c\tau}$ and increases more rapidly
when $1-e^{-c\tau}< p \le 1$.
This is clearly depicted in Fig.\ \ref{figr}.
This result suggests that the decoherence factor ${\rm Re}~r_+$
is small for a large dissipation rate $p>1-e^{-c\tau}$.
In addition, ${\rm Re}~e^{-r_-}$ rapidly converges to 0 for
$p>1-e^{-c\tau}$. Thus the dominant term for such a large $p$ is
$b\frac{-r_-}{r_+-r_-}e^{-r_+ t}$, which converges to $b$.
This phenomenon can be understood physically: the environmental system
is wiped out quickly before absorbing coherence information of the
principal system as $p$ approaches to unity. Therefore this effect
may be called {\it quantum wipe effect}. 

\begin{figure}[hpbt]\begin{center}
\scalebox{0.6}{\includegraphics{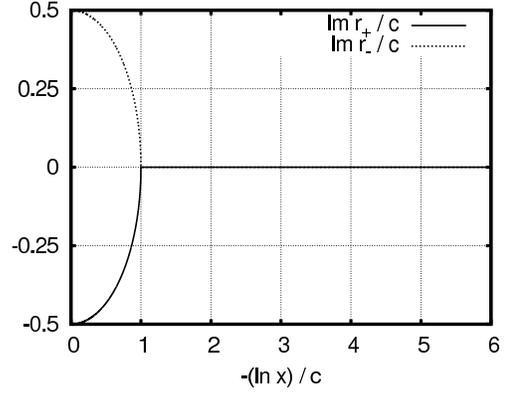}}
\caption{\label{figi}Plot of ${\rm Im}~r_\pm/c$ as functions of 
$-(\ln x) / c$. These imaginary factors vanish for $-\ln x \ge c$
(i.e., $p\ge 1-e^{-c\tau}$).}
\end{center}\end{figure}

In addition, the behaviour of ${\rm Im}~r_\pm$ is rather simple, as
shown in Fig.\ \ref{figi}. They vanish for $p\ge 1-e^{-c\tau}$; this
suggests that we do not observe oscillation in $\eta(t)$ for such a large
$p$.

\begin{figure}[bpt]
\begin{center}
 \scalebox{0.6}{\includegraphics{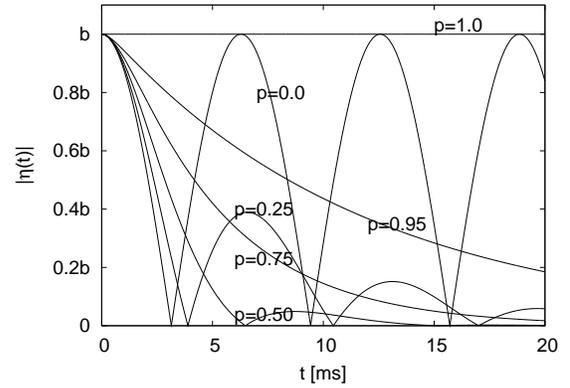}}\caption{\label{figexample}
Time evolution of $|\eta(t)|$ for several different values of $p$
($0.0$, $0.25$, $0.5$, $0.75$, $0.95$, and $1.0$) when
$\tau=1.0\times 10^{-3}$s and $c=1.0\times 10^3$Hz.
}
\end{center}
\end{figure}

As an example, we will see the time evolution of $|\eta(t)|$
for $\tau=1.0\times 10^{-3}$s and $c=1.0\times 10^3$Hz.
Figure \ref{figexample} shows clear suppression of decoherence for
small $p$ and also for large $p$. Suppression of oscillation
is found for large $p$.

The phenomenon we have seen above can be found also in
a spin-boson linear coupling model.
Let us consider a system consisting of a single spin-$1/2$ (system $1$)
and a mode of bosons (system $2$) coupled with the spin system. We
assume that the mode of the bosons is in resonance with the precession
of the spin system (the resonance frequency is $\nu$). Assuming that
systems $1$ and $2$ are surrounded by a bath of bosons of a cavity
and/or a sample holder, we employ the same model illustrated in
Fig.\ \ref{figsimplemodel}.

Let us set the initial state of the spin to
\[
 \rho^{[1]}(0)=\begin{pmatrix}1/2&1/2\\1/2&1/2\end{pmatrix}.
\]
The initial state of the bosonic system is
$
 \rho^{[2]}(0)=e^{-\beta H^{[2]}}/Z
$
where $\beta$ is defined as $\beta=(k_{\rm B}T)^{-1}$
($k_{\rm B}$ is the Boltzmann constant and $T$ is temperature),
$H^{[2]}=\nu a^\dagger a$ is the bosonic system Hamiltonian
($a^\dagger$ and $a$ are the creation and annihilation operators),
and $Z$ is the partition function. The thermal state probabilistically
replacing $\rho^{[2]}$ is $\sigma=\rho^{[2]}(0)$. The initial state of
the total system is set to
$
 \rho^{[1,2]}(0)=\rho^{[1]}(0)\otimes \rho^{[2]}(0).
$
Let the full Hamiltonian be
$H=H^{[1]}+H^{[2]}+H_c$ with the spin-system Hamiltonian
$H^{[1]}=\nu I_z$
and the coupling Hamiltonian $H_c=c I_z(a^\dagger + a)$
(here, $c$ is a coupling constant).

A numerical simulation is used to compute the time evolution governed
by Eq.\ (\ref{eqMain}). We have taken the parameters:
$\nu=3.4\times10^{10}$Hz, $T=1.0$mK,
$c=1.0\times 10^7$Hz, and
$\tau=1.0\times 10^{-8}$s.
The values of $\nu$ and $T$ are taken from physically available values
for an electron spin coupled with an on-resonance bosonic mode (a cavity
mode and/or a phonon mode of a solid sample) in a low-temperature
Q-band electron-nuclear double resonance (Q-band ENDOR) system. In
simulations, $\DD t$ is set to $5.0\times 10^{-10}$s.

\begin{figure}[hpbt]
\begin{center}
 \scalebox{0.6}{\includegraphics{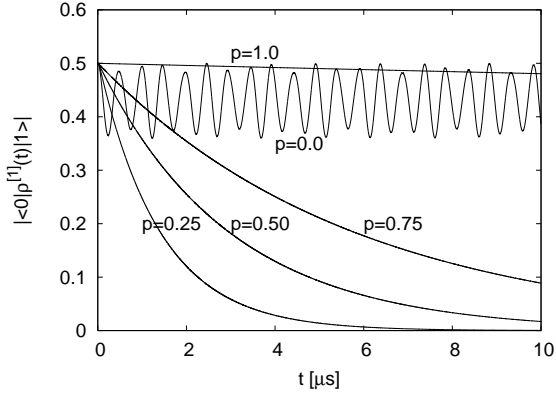}}
\caption{\label{figSB}Numerically computed time evolution
of $|\langle 0|\rho^{[1]}(t)|1\rangle|$ for several different values of
$p$.}
\end{center}
\end{figure}

The simulation showed a clear suppression of decoherence in
$|\langle 0|\rho^{[1]}(t)|1\rangle|=
|\langle 0|{\rm Tr}_2\rho^{[1,2]}(t)|1\rangle|$ for large $p$
as shown in Fig.\ \ref{figSB}.

For the next example, we numerically evaluate the decay of entanglement of
a principal system (system 1) consisting of one electron spin and one
nuclear spin each of which is coupled with an on-resonance bosonic mode
(system 2). The measure of entanglement that we use here is
negativity \cite{Z98,VW02}
\footnote{Negativity $\mathcal{N}$ can be calculated for a density matrix
$\rho^{[a,b]}$ of a bipartite system consisting of subsystems $a$ and
$b$ in the following way:
$
 \mathcal{N}(\rho^{[a,b]})=
[\|(I\otimes\Lambda_\mathrm{T})\rho^{[a,b]}\|-1]/2
$
where $\Lambda_\mathrm{T}$ is the transposition map acting on the
subsystem $b$ and hence $I\otimes\Lambda_\mathrm{T}$ is a partial
transposition map.}.
The evolution of the systems is governed by Eq.\ (\ref{eqMain}).
Let us consider the following Hamiltonians (the subscripts $0$ and $1$
denote labels assigned to individual spins and corresponding
on-resonance bosonic modes).
(i) The Hamiltonian of the principal system:
$H^{[1]}=\nu_0 I_{z0}+\nu_1 I_{z1}+A_{01}I_{z0}I_{z1}$
with the precession frequencies
$\nu_0,\nu_1$ and the spin-spin coupling constant $A_{01}$ (here,
$I_{zi}$ is an $I_z$ operator acting on the $i$th spin).
(ii) The Hamiltonian of the bosonic system:
$H^{[2]}=\nu_0a_0^\dagger a_0+\nu_1a_1^\dagger a_1$.
(iii) The Hamiltonian of spin-boson couplings:
$H_c=c_0I_{z0}(a_0^\dagger+a_0)+c_1I_{z1}(a_1^\dagger+a_1)$ with constants
$c_0$ and $c_1$.
The total Hamiltonian is $H=H^{[1]}+H^{[2]}+H_c$.

The initial reduced density matrix of the principal system is set to
\[
\rho^{[1]}(0)=\frac{|00\rangle\langle00|+|00\rangle\langle11|
+|11\rangle\langle00|+|11\rangle\langle11|}{2}
\]
and that of the bosonic system is set to the thermal one
$\rho^{[2]}(0)=\sigma=e^{-\beta H^{[2]}}/Z$. We set
$
 \rho^{[1,2]}(0)=\rho^{[1]}(0)\otimes \rho^{[2]}(0).
$
The following constants are used:
$\nu_0=3.4\times 10^{10}$Hz, $\nu_1=4.87\times 10^7$Hz,
$A_{01}=1.0\times 10^7$Hz, $T=1.0\times 10^{-3}$mK,
$c_0=c_1=1.0\times 10^7$Hz, and $\tau=1.0\times 10^{-8}$s.
Here, the values of $\nu_0$, $\nu_1$, and $T$ are chosen by considering an
electron spin and a nuclear spin coupled with on-resonance bosonic modes
in a Q-band ENDOR system.
A numerical simulation is performed with $\DD t = 5.0\times 10^{-10}$s.

\begin{figure}[hpbt]
\begin{center}
\scalebox{0.6}{\includegraphics{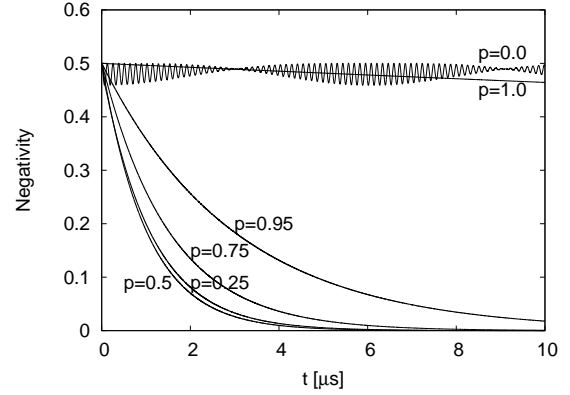}}
\caption{\label{figneg}Numerically computed
time evolution of negativity of the principal system for several values of $p$.}
\end{center}
\end{figure}

We show the time evolution of negativity of the principal system
in Fig.\ \ref{figneg}. An improvement of the entanglement
conservation is found for large $p$ and it is especially conspicuous
for $p\ge0.95$. 

In summary, we considered a model in which a principal system is coupled
with an environmental system that is probabilistically replaced with a
thermal environmental system. In such a model, a high dissipation rate
of the environmental system results in a conservation of coherence of
the principal system. This phenomenon may be called quantum wipe effect.
We have shown that this phenomenon is found in a simple model of
qubit-qubit coupling by an analytical calculation. An increase of a
dissipation rate over a certain threshold value results in coherence
conservation while an increase of the dissipation rate otherwise results
in larger decoherence, as shown in Fig.\ \ref{figr}. This is reminiscent
of the relation between the Zeno effect and the Anti-Zeno effect
\cite{KK00}, namely that very quick applications of projective
operations are effective for coherence conservation while slow
applications of them result in larger decoherence. The quantum wipe
effect has been also found numerically for a model of spin-boson linear
coupling.

The usefulness of the quantum wipe effect is dependent on
how small the threshold value of the dissipation rate is. A coupling
constant is required to be small to obtain a small threshold value. It
is thus expected that this effect will be verified experimentally in
future with a weak system-environment coupling.

AS is supported by the JSPS Research Fellowship for Young Scientists.
RR is supported by the Sasakawa Scientific Research Grant from JSS.
MN would like to thank for a partial support of the Grant-in-Aid
for Scientific Research from JSPS (Grant No. 19540422).
This work is partially supported by ``Open Research Center'' Project
for Private Universities: matching fund subsidy from MEXT.

\end{document}